\documentclass[11pt]{article}
\usepackage{amsmath}
\usepackage{amsthm}
\usepackage{titlesec}
\usepackage{indentfirst}
\theoremstyle{definition}\newtheorem{Df}{Definition}
\theoremstyle{plain}\newtheorem{Th}{Theorem}
\theoremstyle{definition}\newtheorem{Rm}{Remark}
\theoremstyle{definition}
\theoremstyle{plain}\newtheorem{Pp}[Th]{Proposition}
\theoremstyle{plain}
\theoremstyle{plain}
\begin{document}
\title{{\bf A Polynomial-Time Algorithm for the Equivalence between Quantum Sequential Machines\thanks{This research is
supported in part by the National Natural Science Foundation (No.
90303024, 60573006), the Higher School Doctoral Subject Foundation
of Ministry of Education (No. 20050558015), and the Natural
Science Foundation of Guangdong Province (No. 020146, 031541) of
China.}}}
\author{Lvzhou Li,\hskip 2mm Daowen Qiu\thanks{Corresponding author.  {\it E-mail
address:}
issqdw@mail.sysu.edu.cn (D. Qiu).} \\
\small{{\it Department of
Computer Science, Zhongshan University, Guangzhou 510275,}}\\
\small {{\it People's Republic of China}}}
\date{ }
\maketitle

 \vskip 2mm \noindent {\bf Abstract}
\par
\vskip 1mm \hskip 2mm {\it Quantum sequential machines} (QSMs) are
a quantum version of {\it stochastic sequential machines} (SSMs).
Recently, we showed that two QSMs ${\cal M}_{1}$ and ${\cal
M}_{2}$ with $n_{1}$ and $n_{2}$ states, respectively, are
equivalent iff they are $(n_{1}+n_{2})^{2}$--equivalent
(Theoretical Computer Science 358 (2006) 65-74). However, using
this result to check the equivalence likely needs exponential
expected time. In this paper, we consider the time complexity of
deciding the equivalence between QSMs and related problems. The
main results are as follows: (1) We present a polynomial-time
algorithm for deciding the equivalence between QSMs, and, if two
QSMs are not equivalent, this algorithm will produce an
input-output pair with length not more than $(n_{1}+n_{2})^{2}$.
(2) We improve the bound for the equivalence between QSMs from
$(n_1+n_2)^2$ to $n_1^2+n_2^2-1$, by employing Moore and
Crutchfield's method (Theoretical Computer Science
 237 (2000) 275-306). (3) We give that two MO-1QFAs
with $n_{1}$ and $n_{2}$ states, respectively, are equivalent iff
they are $(n_{1}+n_{2})^{2}$--equivalent, and further obtain a
polynomial-time algorithm for deciding the equivalence between two
MO-1QFAs. (4) We provide a counterexample showing that  Koshiba's
method to solve the problem of deciding  the equivalence between
MM-1QFAs may be not valid, and thus the problem is left open
again.


\vskip 2mm

\section*{1. Introduction}

Over the past two decades, quantum computing has attracted wide
attention in the academic community \cite{21,29}. To a certain
extent, this was motivated by the exponential speed-up of Shor's
quantum algorithm for factoring integers in polynomial time
\cite{33} and afterwards Grover's algorithm of searching in
database of size $n$ with only $O(\sqrt{n})$ accesses \cite{19}.

Quantum computers---the physical devices complying with the rules
of quantum mechanics were first considered by Benioff \cite{5},
and then suggested by Feynman \cite{15}. By elaborating and
formalizing Benioff and Feynman's idea, in 1985, Deutsch \cite{13}
re-examined the Church-Turing Principle and defined {\it quantum
Turing machines} (QTMs). Subsequently, Deutsch \cite{14}
considered quantum network models. In 1993, Yao \cite{35}
demonstrated the equivalence between QTMs and quantum circuits.
Quantum computation from the viewpoint of complexity theory was
first studied systematically by Bernstein and Vazirani \cite{12}.

Another kind of simpler models of quantum computation is {\it
quantum finite automata} (QFAs), that can be thought of as
theoretical models of quantum computers with finite memory. This
kind of computing machines was firstly studied independently by
Moore and Crutchfield \cite{28}, as well as Kondacs and Watrous
\cite{26}. Then it was deeply dealt with by Ambainis and Freivalds
\cite{1}, Brodsky and Pippenger \cite{11}, and the other authors
(e.g., name only a few, [2-4,7-10,18], and for the details we may
refer to \cite{21}). The study of QFAs is mainly divided into two
ways: one is {\it one-way quantum finite automata} (1QFAs) whose
tape heads only move one cell to right at each evolution, and the
other is {\it two-way quantum finite automata} (2QFAs), in which
the tape heads are allowed to move towards right or left, or to be
stationary. (Notably, Amano and Iwama \cite{3} dealt with an
intermediate form called 1.5QFAs, whose tape heads are allowed to
move right or to be stationary, and, particularly, they showed
that the emptiness problem for this restricted model is
undecidable.) Furthermore, by means of the measurement times in a
computation, 1QFAs have two fashions: {\it measure-once} 1QFAs
(MO-1QFAs) proposed by Moore and Crutchfield \cite{28}, and, {\it
measure-many} 1QFAs (MM-1QFAs) studied first by Kondacs and
Watrous \cite{26}.

The characteristics of quantum principles can essentially
strengthen the power of some models of quantum computing, but the
unitarity and linearity of quantum physics also lead to some
weaknesses. We briefly state some essential differences between
the QFAs stated above and their classical counterparts by two
aspects. One is from their power. The class of languages
recognized by MM-1QFAs with bounded error probabilities is
strictly bigger than that by MO-1QFAs, but both MO-1QFAs and
MM-1QFAs recognize proper subclass of regular languages with
bounded error probabilities [10,11,26,28]. (Also, the class of
languages recognized by MM-1QFAs with bounded error probabilities
is not closed under the binary Boolean operations \cite{11,10}.)
Kondacs and Watrous \cite{26} proved that some 2QFA can recognize
non-regular language $L_{eq}=\{a^{n}b^{n}|n>0\}$ with one-sided
error probability in linear time (Freivalds \cite{16} proved that
{\it two-way probabilistic finite automata} can recognize
non-regular language $L_{eq}$ with arbitrarily small error, but it
requires exponential expected time \cite{23}. As it is well-known,
classical two-way finite automata can accept only regular
languages \cite{24}).

The other difference is from the viewpoint of decidability. By
$P_{A}(x)$ we denote the probability of the automaton $A$
accepting input string $x$. Then the four cut-point languages,
recognized by $A$ with cut-point $\lambda\in [0,1]$, are defined
by $L_{\bowtie}=\{x:P_{A}(x)\bowtie\lambda\}$, for $\bowtie\in
\{<,\leq,>,\geq\}$. When $A$ is an MM-1QFA, Blondel {\it et al.}
\cite{9} proved that the problems of determining whether
$L_{\bowtie}$ $(\bowtie\in \{<,>\})$ are empty are decidable, but
when $\bowtie\in \{\leq,\geq\}$, such problems are undecidable. In
contrast, when $A$ is a probabilistic automaton, all these
emptiness problems for $\bowtie\in \{<,\leq,>,\geq\}$ are
undecidable \cite{30}.

Recently, another finding is concerning an essential difference
between {\it quantum sequential machines} (QSMs) and {\it
stochastic sequential machines} (SSMs). SSMs \cite{30} may be
viewed as a generalization of probabilistic automata [32,30],
since an SSM that has only one output element and some accepting
states are assigned, reduces to a probabilistic automaton. Two
SSMs, say ${\cal M}_{1}$ and ${\cal M}_{2}$ having $n_{1}$ and
$n_{2}$ states, respectively, and the same input and output
alphabets, are called equivalent if they have equal accepting
probability for any input-output pair $(u,v)$. As was known, a
crucial result concerning SSMs by Paz \cite{30} is that ${\cal
M}_{1}$ and ${\cal M}_{2}$ are equivalent iff they are
$(n_{1}+n_{2}-1)$--equivalent (that is, their accepting
probabilities are equal for any input-output pair whose length is
not more than $n_{1}+n_{2}-1$). Recently, Gudder \cite{20} first
defined {\it sequential quantum machines} (SQMs), a quantum
analogue of SSMs, and Gudder asked whether or not such an
equivalence consequence also holds for SQMs. Then, Qiu \cite{31}
re-defined an equivalent version called {\it quantum sequential
machines} (QSMs), that were formally a quantum counterpart of
SSMs, just as {\it quantum finite automata} (QFAs) to
probabilistic automata.  Qiu \cite{31} demonstrated that there are
two QSMs with $n_{1}$ and $n_{2}$ states, respectively, such that
they are $(n_{1}+n_{2}-1)$--equivalent, but {\it not} equivalent,
after all. Hence, Gudder's problem was given a negative answer.

Latterly, we \cite{27} further proved that two QSMs ${\cal M}_{1}$
and ${\cal M}_{2}$ having $n_{1}$ and $n_{2}$ states,
respectively, and the same input and output alphabets $I$ and $O$,
are equivalent iff they are $(n_{1}+n_{2})^{2}$--equivalent, a new
feature in contrast to the $(n_{1}+n_{2}-1)$--equivalence for SSMs
\cite{30}. However, using this result to check the equivalence
between QSMs needs exponential expected time
$(O(m^{(n_{1}+n_{2})^{2}}))$ where $m=|I|\times |O|$.

The remainder of the paper is organized as follows. In Section 2,
we recall the definitions of SSMs and QSMs, and related results.
Section 3 is the main part. In Subsection 3.1, we detail a
polynomial-time algorithm ($O(m.(n_{1}+n_{2})^{12}$)) for deciding
the equivalence between QSMs. In Subsection 3.2, we improve the
bound for the equivalence between QSMs from $(n_1+n_2)^2$ to
$n_1^2+n_2^2-1$, by employing Moore and Crutchfield's method
\cite{28}. Section 4 is concerning the equivalence between
\textit{one-way} QFAs. In Subsection 4.1, we provide a
polynomial-time algorithm for deciding the equivalence between
MO-1QFAs. In Subsection 4.2, we provide a counterexample showing
that the method stated in \cite{25} to decide the equivalence
between MM-1QFAs is not valid. Finally, some remarks are made in
Section 5 to conclude this paper.

\section*{2. Preliminaries}

In this section, we briefly review some definitions and related
properties that will be used in the sequel.

Firstly, we explain some notations. An $n$-dimensional row vector
$(a_{1}\hskip 1mm a_{2} \dots a_{n})$ is called {\it stochastic}
if $a_{i}\geq 0$ $(i=1,2,\ldots,n)$, and $\sum_{i=1}^{n}a_{i}=1$;
in particular, it is called a {\it degenerate stochastic vector}
if it has 1 only in one entry and else $0$s.  A matrix is called
{\it stochastic} if its each row is a stochastic vector. As usual,
for non-empty set $I$, by $|I|$ we mean the cardinality of $I$,
and by $I^{*}$ we mean the set of all finite length strings over
$I$. For $u\in I^{*}$, $|u|$ denotes the length of $u$; when
$|u|=0$, $u$ is an empty string, denoted by $\epsilon$. We denote
$I^{+}=I^{*}-\{\epsilon\}$. For input alphabet $I$ and output
alphabet $O$, the set of all {\it input-output} pairs is defined
as
\[
\{(u,v)\in I^{*}\times O^{*}:|u|=|v|\}.
\]
For any input-output pair $(u,v)$,  we denote by $l(u,v)$  the
length of $u$ or $v$.

For the details on  {\it stochastic sequential machines} (SSMs),
we can refer to [30]. Next, we recall the definition of {\it
quantum sequential machines} (QSMs), a quantum counterpart of SSMs
[30].
\begin{Df}[{[31]}]
 A QSM is a 5-tuple
 ${\cal M}=(S,\eta_{i_0},I,O,\{A(y|x): x\in I, y\in O\})$ where
 $S=\{s_1,s_2,\dots,s_n\}$ is a finite set of internal
 states; $\eta_{i_0}$ is an $n-$dimensional degenerate stochastic row
 vector; $I$ and
 $O$ are input and output alphabets, respectively; $A(y|x)$ is an
 $n\times n$ complex matrix satisfying
 $\sum_yA(y|x)A(y|x)^\dagger=\bf I
 $
 for any
 $x\in I$, where the symbol $\dagger$ denotes Hermitian conjugate
 operation and $\bf I$ is unit matrix. In particular, for
input-output pair $(\epsilon,\epsilon)$, $A(\epsilon|\epsilon)=\bf
I$.
\end{Df}
\begin{Rm}
It is worth pointing out that, before the definition of QSMs [31],
Gudder [20] first defined \textit{sequential quantum machines}
(SQMs), an equivalent version of QSMs. The equivalence between
QSMs and SQMs was proved by Qiu [31]. The reader can refer to
[20,31] for more information.
\end{Rm}

In brief, we may denote QSM ${\cal M}$ as
$(S,\eta_{i_0},I,O,\{A(y|x)\})$. In the QSM ${\cal M}$ defined
above, if matrix $A(y|x)=[a_{ij}(y|x)]$, then $a_{ij}(y|x)$ (resp.
$|a_{ij}(y|x)|^{2}$) represents the amplitude (resp. the
probability) of the machine entering state $s_{j}$ and yielding
$y$ after $x$ being inputted with the present state $s_{i}$. Thus,
given the QSM ${\cal M}$ above, we let $P_{\cal M}(v|u)$ denote
the probability of ${\cal M}$ printing the word $v$ after having
been fed with the word $u$, and it is defined as follows:
\begin{align}
P_{\cal M}(v|u)=\begin{Vmatrix}\eta_{i_0}A(v|u)\end{Vmatrix}^2,
\end{align}
where $\eta_{i_0}$ is the initiation-state distribution of ${\cal
M}$, $(u,v)=(x_1x_2\ldots x_m,y_1y_2\ldots y_m)$ denotes an
input-output pair, and $A(v|u)=A(y_1|x_1)A(y_2|x_2)\cdots
A(y_m|x_m)$. Clearly, we have
\begin{align}
P_{\cal
M}(v|u)=\eta_{i_0}A(v|u)A(v|u)^{\dagger}\eta_{i_0}^{\dagger}.
\end{align}

If the initiation-state distribution $\eta_{i_0}$ in QSM ${\cal
M}$ is omitted, then we call ${\cal M}$ {\it uninitiated QSM}
(UQSM). For a UQSM ${\cal M}$, by $P_{\cal M}^{\eta_{i_0}}(v|u)$
we mean the probability that, with the initiation-state
distribution $\eta_{i_0}$ being specified, ${\cal M}$ prints $v$
after $u$ being inputted.

\begin{Df} Let ${\cal M}$ be a UQSM. Any two initiation-state distributions $\eta_{i_0}$ and
$\eta_{j_0}$ of ${\cal M}$ are said to be equivalent (resp.
$k$--equivalent) with respect to ${\cal M}$, if $P_{\cal
M}^{\eta_{i_0}}(v|u)=P_{\cal M}^{\eta_{j_0}}(v|u)$ for any
input-output pair $(u,v)$ (resp. for any input-output pair $(u,v)$
with $l(u,v)\leq k $).
\end{Df}

In the following, we define the equivalence between machines.
\begin{Df}
Two machines (SSMs, SQMs, or QSMs) ${\cal M}_1$ and ${\cal M}_2$
with the same input and output alphabets are called equivalent
(resp. $k$--equivalent) if $P_{{\cal M}_1}(v|u)=P_{{\cal
M}_2}(v|u)$ for any input-output pair $(u,v)$ (resp. for any
input-output pair $(u,v)$ with $l(u,v)\leq k $).
\end{Df}
\begin{Rm}
Given a QSM ${\cal M}$, from Eq. (1), we know $P_{{\cal
M}}(\epsilon|\epsilon)\equiv 1$. Therefore, in Definition 3, we
can require $l(u,v)\geq 1$. In what follows, for verifying the
equivalence between QSMs, we only consider the input-output pair
$(u,v)$ with $l(u,v)\geq 1$.
\end{Rm}

A crucial result concerning SSMs is that two SSMs with $n_{1}$ and
$n_{2}$ states, respectively, are equivalent iff they are
$(n_{1}+n_{2}-1)$--equivalent [30]. Therefore, Gudder asked
whether it holds for SQMs. Then Qiu [31] gave its negative answer.
Recently, we further showed that two QSMs  ${\cal M}_{1}$ and
${\cal M}_{2}$ are equivalent iff they are
$(n_{1}+n_{2})^{2}$--equivalent [27]. The two results are stated
in the following theorem.

\begin{Th}[{[31,27]}]
(1) There exist QSMs (or SQMs) ${\cal M}_1$ and ${\cal M}_2$ with
$n_{1}$ and $n_{2}$ states, respectively, and the same input and
output alphabets, such that though ${\cal M}_1$ and ${\cal M}_2$
are $(n_{1}+n_{2}-1)$--equivalent, they are {\it not} equivalent.

(2) Two machines (SQMs or QSMs) ${\cal M}_1$ and ${\cal M}_2$ with
$n_{1}$ and $n_{2}$ states, respectively, and the same input and
output alphabets, are equivalent iff they are
$(n_{1}+n_{2})^{2}$--equivalent.

\end{Th}

\section*{3. Equivalence between QSMs}
 In this section, we consider further the equivalence between QSMs. In Subsection 3.1, based on the way stated in [27],
 we give a polynomial-time algorithm that
 takes as input two QSMs and determines whether they are equivalent.
 In Subsection 3.2, by employing Moore and Crutchfield's method [28], we give a better bound for the
equivalence between QSMs.

\subsection*{ 3.1. A polynomial-time algorithm for  the equivalence
between QSMs}

As stated before, directly testing Theorem 1 (2) for the
equivalence between QSMs needs exponential expected time.
Therefore, in this subsection, we present a polynomial-time
algorithm for the equivalence between QSMs.

Before presenting the algorithm,  we recall the definition of
direct sum of two matrices. Suppose that $A_{mn}$ and $B_{kl}$ are
$m\times n$ and $k\times l$  matrices, respectively. Then the
direct sum $A_{mn}\oplus B_{kl}$ is an $(m+k)\times (n+l)$ matrix,
defined as $A_{mn}\oplus B_{kl}=\begin{bmatrix}
  A_{mn} & 0 \\
  0 & B_{kl}
\end{bmatrix}.$

Now, we can present the main theorem as follows.

\begin{Th}
There is a polynomial-time algorithm  that takes as input two QSMs
${\cal M}_1$ and ${\cal M}_2$ and determines whether ${\cal M}_1$
and ${\cal M}_2$ are equivalent. Furthermore, if the two QSMs are
not equivalent, then the algorithm outputs an input-output pair
$(u,v)$ satisfying that  $P_{{\cal M}_{1}}(v|u)\not=P_{{\cal
M}_{2}}(v|u)$, and $l(u,v)\leq (n_1+n_2)^2$, where $n_1$ and $n_2$
are the numbers of states in ${\cal M}_1$ and ${\cal M}_2$,
respectively.
\end{Th}

\noindent{\bf Proof.} Given two QSMs having the same input and
output alphabets:
\begin{align*}{\cal
M}_{1}=(S_{1},\eta_{i_0},I,O,\{A_{1}(y|x)\})
\hspace{2mm}\text{and}\hspace{2mm} {\cal
M}_{2}=(S_{2},\eta_{j_0},I,O,\{A_{2}(y|x)\}),
\end{align*}
 where
$|S_{1}|=n_{1}$ and $|S_{2}|=n_{2}$. We construct UQSM  ${\cal
M}=(S, I, O, \{A(y|x):x\in I,y\in O\})$, where $S=S_{1}\cup
S_{2}$, $A(y|x)=A_{1}(y|x)\oplus A_{2}(y|x)$. For any input-output
pair $(u,v)$, denote
 \begin{equation}
D(v|u)= A(v|u)A(v|u)^{\dagger}. \end{equation} Then for any
input-output pairs $(u,v)$ and $(x,y)$ where $l(x,y)=1$, we have
\begin{eqnarray}
D(yv|xu)&=&\nonumber A(yv|xu)A(yv|xu)^{\dagger}\\
&=& \nonumber A(y|x)A(v|u)(A(y|x)A(v|u))^{\dagger}\\
&=& A(y|x)D(v|u)A(y|x)^{\dagger}.
\end{eqnarray}
Let
\begin{equation}
{\cal D}=\{ D(v|u): (u,v)\hskip 1mm {\rm is} \hskip 1mm {\rm
input-output}\hskip 1mm {\rm pair}, \hskip 1mm {\rm and } \hskip
1mm l(v,u)\geq 1\}, \end{equation} and $\rho=(\eta_{i_0},\bf{0})$
and $\rho^{'}=(\bf{0},\eta_{j_0})$, where $\rho$ and $\rho^{'}$
are $(n_1+n_2)$-dimensional row vectors, and can serve as two
different initiation-state distributions of ${\cal M}$. Then for
any input-output pair $(u,v)$, we have
\begin{eqnarray}
P_{{\cal M}}^{\rho}(v|u)&=&\nonumber \rho
A(v|u)A(v|u)^{\dagger}\rho^{\dagger}\\
&=&\nonumber \eta_{i_0}A_{1}(v|u)A_{1}(v|u)^{\dagger}\eta_{i_0}^{\dagger}\\
&=&P_{{\cal M}_{1}}(v|u),
\end{eqnarray}
and similarly,
\begin{equation}
P_{{\cal M}}^{\rho^{'}}(v|u)=P_{{\cal M}_{2}}(v|u).
\end{equation}
Therefore, ${\cal M}_1$ and ${\cal M}_2$ are equivalent (i.e.,
$P_{{\cal M}_{1}}(v|u)= P_{{\cal M}_{2}}(v|u)$ for any
input-output pair $(u,v)$) if and only if  $\rho$ and $\rho^{'}$
are equivalent with regard to ${{\cal M}}$, that is, for any
$D(v|u)\in {\cal D}$,
\begin{align}
 \hspace{2mm}\rho
D(v|u)\rho^\dagger=\rho^{'} D(v|u){\rho^{'}}^\dagger.
\end{align}

Let $\Phi({\cal D})$ be the linear subspace spanned by ${\cal D}$,
and let ${\cal B}$ be a basis for the subspace $\Phi({\cal D})$.
Clearly, the total space as to $\Phi({\cal D})$ consists of all
$(n_1+n_2)-$order complex square matrices, together with the usual
operations of matrices, whose dimension is $(n_1+n_2)^2$. Hence
${\cal B}$ has at most $(n_1+n_2)^2$ elements, and the two QSMs
${\cal M}_1$ and ${\cal M}_2$ are equivalent if and only if Eq.
(8) holds for every vector $D(v|u)\in {\cal B}$.

\paragraph{Design of the algorithm.} Without loss of generality, we assume that $I=\{a\}$ and
$O=\{0,1\}$. Then we define binary tree $T$ as follows. Tree $T$
has a corresponding node $D(v|u)$ (defined in Eq. (3)) for every
input-output pair $(u,v)\in I^{*}\times O^{*}$. The root of $T$ is
$D(\epsilon|\epsilon)$ that is identity matrix $I$. In the
definition of ${\cal D}$ (Eq. (5)), we notice that $l(v,u)\geq 1$,
so, we exclude $D(\epsilon|\epsilon)$ when searching for the basis
${\cal B}$ of subspace $\Phi({\cal D})$. However, we still make it
act as the root of tree $T$ for the sake of convenience. Every
node $D(v|u)$ in $T$ has two children $D(0v|au)$ and $D(1v|au)$.
For any $x\in I$, and $y\in O$, $D(yv|xu)$ can be calculated from
its parent $D(v|u)$ by Eq. (4).

Our algorithm is described in Figure 1 which is to efficiently
search for the basis ${\cal B}$ of $\Phi({\cal D})$ by pruning
tree $T$. In the algorithm, $queue$ denotes a queue, and ${\cal
B}$, the basis stated above, is initially  set to be the empty
set. We  visit tree $T$ by breadth-first order. At each node
$D(v|u)$, we verify whether it is linearly independent of ${\cal
B}$. If it is, we add  it to ${\cal B}$. Otherwise, we prune the
subtree rooted at $D(v|u)$. We stop traversing tree $T$ after
every node in $T$ has been either visited or pruned. The vectors
in the resulting set ${\cal B}$ will form a basis for $\Phi({\cal
D})$, which will be proven later on. At the end of the algorithm,
we verify whether Eq. (8) holds for every vector in ${\cal B}$. If
yes, then the two QSMs are equivalent. Otherwise, the algorithm
returns an input-output pair $(u,v)$ satisfying that $P_{{\cal
M}_{1}}(v|u)\not=P_{{\cal M}_{2}}(v|u)$.

\begin{center}
 Figure 1.
 Algorithm for the equivalence between QSMs.
 \fbox{\parbox{16cm}{
 {\small
\begin{quote} {\bf Input:} ${\cal
M}_1=(S_1,\eta_{i_0},\{a\},\{0,1\},\{A_{1}(y|x)\})$, and ${\cal
M}_2=(S_2,\eta_{j_0},\{a\},\{0,1\},\{A_{2}(y|x)\})$
\begin{quote}
 Set ${\cal B}$  to be the empty set;\\
 $queue\leftarrow D(\epsilon|\epsilon)$;\\
 {\bf while} $queue$ is not empty {\bf do}\\
 {\bf begin} take an element $D(v|u)$ from $queue$;
 \begin{quote}
{\bf if} $D(v|u) \notin span({\cal B})$ {\bf
then} \\
{\bf begin}  add vector $D(v|u)$ to ${\cal B}$;
//$D(\epsilon|\epsilon)$ is not added to ${\cal B}$.
\begin{quote}
 add  $D(0v|au)$ and $D(1v|au)$
to $queue$;
\end{quote}
{\bf end};
 \end{quote}
 {\bf end};
\end{quote}
\begin{quote}
{\bf if} $\forall B\in {\cal B}$, $(\eta_{i_0},{\bf
0})B(\eta_{i_0}, {\bf 0})^{\dagger}=( {\bf 0},\eta_{j_0})B( {\bf
0},\eta_{j_0})^{\dagger}$
{\bf then} return $(yes)$\\
{\bf else} return (the  pair $(u,v)$: $(\eta_{i_0},{\bf
0})B(\eta_{i_0}, {\bf 0})^{\dagger}\neq( {\bf 0},\eta_{j_0})B(
{\bf 0},\eta_{j_0})^{\dagger}$);
\end{quote}
\end{quote}
}
 }
 }
\end{center}

\begin{Rm}
The basic idea regarding our algorithm is to efficiently search
for the basis ${\cal B}$ of $\Phi({\cal D})$. The foundation of
our algorithm is the {\it breadth-first} search for traversing a
tree. The method used in our algorithm is to prune tree $T$. By
pruning some unwanted subtrees, we do not need to visit all nodes
with height not more than $(n_1+n_2)^2$ (here the height of root
is defined as 0), such that we can greatly reduce the number of
nodes to be visited.
\end{Rm}

\paragraph{Validity of the algorithm.}
 Now we explain why the resulting set ${\cal
B}$ form a basis for $\Phi({\cal D})$.  From the analysis above,
we know that after running the algorithm, we will get a pruned
tree. We denote the resulting tree by $T_P$ which is formed by the
nodes in the following set
\begin{align*}
{\cal B}\cup\{D(\sigma_ov|\sigma_iu): D(v|u)\in {\cal B},
D(\sigma_ov|\sigma_iu)\in span({\cal B}), \sigma_i\in I,
\sigma_o\in O\},
\end{align*}
where the former part ${\cal B}$ consists of the internal nodes of
tree $T_P$, and the latter part comprises the leaf nodes of tree
$T_P$.

For $i\geq0$, we let
\begin{align}
{\cal B}_i=\{D(yv|xu): D(v|u)\hspace{1mm} \text{is a leaf in
$T_P$},\hspace{1mm} l(x,y)=i\},
\end{align}
where when $i\geq 1$, set ${\cal B}_i$ consists of  unvisited
nodes in tree $T$ which have distance $i$ from a leaf in tree
$T_P$; when $i= 0$, set ${\cal B}_0$ is the set of  leaves of
$T_P$. Then it can be readily seen that
\begin{align}
{\cal D}={\cal B}\cup\bigcup_{i=0}^{\infty} {\cal B}_i.
\end{align}
Proving that ${\cal B}$ forms a basis for $\Phi({\cal D})$ amounts
to showing that $span({\cal B})\equiv span({\cal D}$).
Equivalently, we only need to prove the following proposition.
\begin{Pp}
For all $i\geq 0$, ${\cal B}_i\subseteq span ({\cal B})$.
\end{Pp}
\noindent\textbf{Proof.} Let ${\cal B}=\{B_1, B_2,\dots, B_m\}$
for some $m\leq (n_1+n_2)^2$. We show the proposition by induction
on $i$. The basic case ${\cal B}_0\subseteq span({\cal B})$
follows straightforward from our analysis above. Now assume that
${\cal B}_i\subseteq span({\cal B})$.  Then for any input-output
pairs $(u,v),(x,y)$, and,  $(\sigma_o, \sigma_I)$, where
$l(\sigma_o, \sigma_I)=1$,  such that $D(v|u)$ is a leaf and
$l(x,y)=i$,  by Eq. (9) we know $D (yv|xu)\in {\cal B}_{i}$, and,
with the assumption, $D (yv|xu)=\sum _{j=1}^{m}\alpha_j B_j$ for
some $\alpha_{j}\in {\bf C}$ $(j=1,2,\ldots,m)$; furthermore, for
any $B_{j}\in {\cal B}$, say $B_{j}=D(v_{j}|u_{j})$ for some
input-output pair $(u_{j},v_{j})$, we have $A(\sigma_o|\sigma_I)
B_j A(\sigma_o|\sigma_I)^\dagger=D(\sigma_ov_{j}|\sigma_Iu_{j})\in
span({\cal B}\cup{\cal B}_0)$. Therefore, we get that
\begin{align*}
D(\sigma_oyv|\sigma_Ixu)&=A(\sigma_o|\sigma_I)D (yv|xu)A
(\sigma_o|\sigma_I)^\dagger\\
&=A(\sigma_o|\sigma_I)\bigg(\sum _{j=1}^{m}\alpha_j
B_j\bigg)A(\sigma_o|\sigma_I)^\dagger\\
&=\sum _{j=1}^{m}\alpha_j\bigg(A(\sigma_o|\sigma_I)
B_j A(\sigma_o|\sigma_I)^\dagger\bigg)\\
&\in span({\cal B}\cup{\cal B}_0)\equiv span({\cal B}).
\end{align*}
This shows that ${\cal B}_{i+1}\subseteq span ({\cal B})$, and the
proposition is proved.
\qed\\

\paragraph{Complexity of the algorithm.}
 Firstly we assume that all the inputs consist of
complex numbers whose real and imaginary parts are rational
numbers and that each arithmetic operation on rational numbers can
be done in constant time. Because the basis ${\cal B}$ has at most
$(n_1+n_2)^2$ elements,  the nodes to be visited will be at most
$O((n_1+n_2)^2)$. (Here we need recall a result that to verify
whether a set of $n$-dimensional vectors is linearly independent
needs time $O(n^{3})$ [17].)  At every visited node $D(v|u)$ the
algorithm may do two things: (i) verifying whether or not the
 $(n_1+n_2)^2$-dimensional vector $D(v|u)$ is linearly
independent of the set ${\cal B}$, which needs time
$O((n_1+n_2)^6)$ according to the result in [17] just stated
above; (ii) calculating its children nodes by Eq. (4) (if
$D(v|u)\notin {\cal B}$), which can be done in time
$O((n_1+n_2)^4)$.
 Thus the total runtime is $O((n_1+n_2)^{12})$.

So far, we have completed the proof of Theorem 2.
\qed\\
\begin{Rm}
In the algorithm above, for convince we consider only the case
where $|I|=1$ and $|O|=2$. In general, let $m=|I|\times |O|$. Then
the algorithm almost keeps on except that  the total nodes to
visit will be at most $O(m.(n_1+n_2)^2)$, and as a result, the
time complexity will be $O(m.(n_1+n_2)^{12})$.
\end{Rm}

\subsection*{3.2. An improved bound for the equivalence between
QSMs} In this subsection, we give an improved bound for the
equivalence between QSMs, using the bilinearization technique
given by Moore and Crutchfield [28]. Firstly, we define a new
model as follows.
\begin{Df}
A bilinear machine (BLM)  is a four-tuple ${\cal M}=( S,
\pi,\{M(\sigma)\}_{\sigma\in\Sigma},\eta)$ over alphabet $\Sigma$,
where $S$ with $|S|=n$ is a finite state set, $\pi\in {\bf
C}^{1\times n}$, $\eta\in {\bf C}^{n\times 1}$ and
$M(\sigma)\in{\bf C}^{n\times n}$ for $\sigma\in \Sigma$.
\end{Df}
Associated to a  BLM ${\cal M}$, the \textit{word function}
$f_{\cal M}:\Sigma^{*}\rightarrow {\bf C}$ is defined in the form:
$f_{\cal M}(w)=\pi M(w_1)\dots M(w_n)\eta$, where $w=w_1\dots
w_n\in \Sigma^{*}$.

A {\it probabilistic automaton} (PA) is a BLM with the restriction
that $\pi$ is a stochastic vector, $\eta$ consists of 0's and 1's
only, and the matrices $M(\sigma)$ $(\sigma\in \Sigma)$ are
stochastic. Then, the word function $f_{\cal M}$ associated to PA
${\cal M}$ has domain in $[0,1]$.
\begin{Df}
Two BLMs (include PAs) ${\cal M}_1$ and ${\cal M}_2$ over the same
alphabet $\Sigma$ are said to be equivalent (resp. $k$-equivalent)
if $f_{{\cal M}_1}(w)=f_{{\cal M}_2}(w)$ for any $w\in \Sigma^{*}
$ (resp. for any input string $w$ with $ |w|\leq k$).
\end{Df}

As stated in Paz [30], the result with regard to the equivalence
between SSMs can also be applied to PAs. Therefore, based on Paz
[30], Tzeng [34] considered further the equivalence between PAs,
giving a polynomial-time algorithm to the problem. Now, the
results are stated in the following.
\begin{Th}[{[30,34]}]
 Two PAs ${\cal M}_1$ and ${\cal M}_2$ with $n_1$ and $n_2$
states, respectively, are equivalent if and only if they are
$(n_1+n_2-1)$-equivalent. Furthermore, there is a polynomial-time
algorithm that takes as input two PAs ${\cal M}_1$ and ${\cal
M}_2$ and determines whether ${\cal M}_1$ and ${\cal M}_2$ are
equivalent.
\end{Th}

\begin{Rm}
In fact, one can  readily find that Paz 's way [30] can also be
applied to BLMs, and the algorithm given by Tzeng [34] still works
for BLMs. Therefore, Theorem 4 holds for the more general
model---BLMs.
\end{Rm}

Next, we transform a QSM to a BLM by the way given by Moore and
Crutchfield [28], and then obtain an improved result for the
equivalence between QSMs. That is the following theorem.
 \begin{Th}
Two QSMs  ${\cal M}_1$ and ${\cal M}_2$ with $n_1$ and $n_2$
states, respectively, are equivalent if and only if they are
$(n_1^2+n_2^2-1)$-equivalent.
 \end{Th}
\noindent\textbf{Proof.} Given an $n$-state QSM ${\cal
M}=(S,\eta_{i_0},I,O,\{A(y|x)\})$, let $h_j$ ($j=1,\dots n$) be a
column vector that has only 1 in the $j$th element and else
$0$s. Then we have\\
\begin{align*}
P_{\cal M}(v|u)&=\begin{Vmatrix}\eta_{i_0}A(v|u)\end{Vmatrix}^2
=\sum^n_{j=1}|\eta_{i_0}A(v|u)h_j|^2\\
&=\sum^n_{j=1}(\eta_{i_0}\otimes\eta_{i_0}^*)\big[A(v|u)\otimes
A(v|u)^*\big](h_j\otimes h_j^*)\\
&=(\eta_{i_0}\otimes\eta_{i_0}^*)\big[A(v|u)\otimes
A(v|u)^*\big]\sum^n_{j=1}(h_j\otimes h_j^*).
\end{align*}
Here we can construct a BLM ${\cal M}^{'}=(S^{'}, \pi,
M(\sigma)_{\sigma\in \Sigma},\eta)$ as follows:
\begin{itemize}
    \item  $|S^{'}|=n^2$, $\pi=\eta_{i_0}\otimes\eta_{i_0}^*$;
    \item   $\Sigma=\{(y|x): y\in O, x\in I\}$, and
    $M((y|x))=A(y|x)\otimes
A(y|x)$;
    \item $\eta=\sum^n_{j=1}(h_j\otimes h_j^*)$.
    \end{itemize}
 Then we get $ P_{\cal M}(y_1\dots y_m|x_1\dots x_m)=f_{{\cal M}^{'}} ((y_1|x_1)\dots (y_m|x_m))$.
 Hence every $n$-state QSM can be transformed to an equivalent $n^2$-state
 BLM, and  by Remark 5, we have proved  the theorem.
\qed\\

\begin{Rm}
Considering the equivalence between two QSMs, we have improved the
bound from $(n_1+n_2)^2$ to $(n_1^2+n_2^2-1)$ by the way of Moore
and Crutchfield [28], which seems to imply that the way used in
Subsection 3.1 is unwanted. Nevertheless, the way used in
Subsection 3.1 offers us a different insight to QSMs and even
other quantum computing models, and maybe can be used to solve
some new problems concerning quantum computing models.
\end{Rm}

\section*{4. Equivalence between one-way QFAs}
In this section, we consider the equivalence between one-way QFAs.
More specifically, in Subsection 4.1, we present a polynomial-time
algorithm for the equivalence between MO-1QFAs, by means of the
idea in Subsection 3.1; in Subsection 4.2, we provide a
counterexample  showing that the method used in [25] to decide the
equivalence between MM-1QFAs may be not valid.

\subsection*{4.1  Equivalence between MO-1QFAs}

  First, we review the definition of
MO-1QFAs [28,11].

An MO-1QFA ${\cal A}$ is a 5-tuple ${\cal A}=(Q,\Sigma,q_{0},
\{A(x):x\in \Sigma \},F)$, where
 $Q$ is a finite set of states (let $|Q|=n$); $q_0$ is the initial
state; $\Sigma$ is a finite set of input symbols;  $A(x)$ denotes
an $n\times n$ unitary evolution matrix for each $x\in \Sigma$;
 $F\subseteq Q$ is the set of accepting states, with
corresponding projection matrix $P_{{\rm acc}}={\rm diag}
(p_{0}\hskip 1mm p_{1}\ldots p_{n-1})$ where for $i=0,\dots ,n-1$,
$p_i$ equals to $1$ if $q_i\in F$ else 0.

 As usual, let $\langle q_{i}|$ denote the $n$-dimensional row
vector $(0\cdots 1\cdots 0)$ whose $(i+1)$th component is $1$ and
the others $0$s ($i=0,1,\ldots,n-1$). Any configuration of ${\cal
A}$ is described by a unit row vector in the {\it superposition}
form $\langle \psi|=\sum_{i=0}^{n-1}\alpha_i\langle q_{i}|$, with
that $\sum_{i=0}^{n-1}|\alpha_{i}|^{2}=1$, and $\alpha_{i}$
denoting the amplitude of ${\cal A}$ being in state $q_{i}$. If
${\cal A}$ is in configuration $\langle \psi|$  and reads an input
symbol $\sigma\in\Sigma$, then the new configuration of ${\cal A}$
becomes $\langle \psi^{'}|=\langle \psi|A(\sigma)$.

 The probability of ${\cal A}$ accepting input
string $u=x_{1}x_{2}\ldots x_{m}$ is defined as
\begin{equation}
P_{{\cal A}}^{q_{0}}(u)=\|\langle q_{0}|A(u)P_{{\rm acc}}\|^{2}
\end{equation}
where $A(u)=A(x_{1})A(x_{2})\cdots A(x_{m})$. For simplicity, we
often write $P_{{\cal A}}^{q_{0}}(u)$ by $P_{{\cal A}}(u)$ if no
confusion results.

We introduce two definitions regarding the equivalence between
MO-1QFAs  as follows.

\begin{Df}
Two MO-1QFAs ${\cal A}_{1}$ and ${\cal A}_{2}$ having the same set
of input symbols are called equivalent (resp. $k$-equivalent) if
for any input string $u$  (resp. for any input string $u$ with $
|u|\leq k$), they have equal accepting probability, i.e.,
$P_{{\cal A}_{1}}(u)=P_{{\cal A}_{2}}(u)$.
\end{Df}

\begin{Df}
Given an MO-1QFAs ${\cal A}$ whose initial state is not specified,
then two states $q_{1}$ and $q_{2}$ in  ${\cal A}$ are called
equivalent (resp. $k$-equivalent) if for any input string $u$
(resp. for any input string $u$ with $ |u|\leq k$), $P_{{\cal
A}}^{q_{1}}(u)=P_{{\cal A}}^{q_{2}}(u)$.
\end{Df}

Concerning the equivalence between MO-1QFAs, we have the following
theorem.

\begin{Th}

Two MO-1QFAs ${\cal A}_{1}$ and ${\cal A}_{2}$ are equivalent if
and only if they are $(n_{1}+n_{2})^2$-equivalent, where $n_{1}$
and $n_{2}$ are the numbers of states in ${\cal A}_{1}$ and ${\cal
A}_{2}$, respectively. Furthermore, there is a polynomial-time
algorithm that takes as input two MO-1QFAs ${\cal A}_1$ and ${\cal
A}_2$ and determines whether ${\cal A}_1$ and ${\cal A}_2$ are
equivalent.

\end{Th}

\noindent\textbf{Proof.} We give a brief proof by  four steps
below.
\begin{enumerate}
\item Let MO-1QFA ${\cal A}=(Q,\Sigma,q_{0}, \{A(x):x\in \Sigma
\},F)$. Then for any $u\in \Sigma^{*}$, we have
\begin{align*}
P_{\cal A}^{q_{0}}(u)&=\begin{Vmatrix}
\langle q_{0}|A(u)P_{{\rm acc}}\end{Vmatrix}^2\\
&=\langle q_{0}|A(u)P_{{\rm acc}}P_{{\rm acc}}^{\dagger}A(u)^{\dagger}|q_{0}\rangle\\
&=\langle q_{0}|A(u)P_{{\rm acc}}A(u)^{\dagger}|q_{0}\rangle,
\end{align*}
where notation $|\cdot\rangle$ denotes the conjugate transpose of
$\langle\cdot|$.

 Denote $F(u)=A(u)P_{{\rm acc}}A(u)^{\dagger}$. Then
\begin{align}
&P_{\cal A}^{q_{0}}(u)=\langle q_{0}|F(u)|q_{0}\rangle,\\
&F(xu)=A(x)F(u)A(x)^{\dagger}.
\end{align}

\item Denote
\begin{align}
&{\cal F}=\{F(u): u\in \Sigma^{*}\},\\
 & {\cal F}(k)=\{F(u): u\in
\Sigma^{*},  |u|\leq k\}.
\end{align}
As we did in the proof of Theorem 1, a set of linearly independent
vectors can be found in ${\cal F}(n^{2})$ $(n=|Q|)$ such that any
vector in ${\cal F}$ is a linearly combination of these vectors.
Therefore, by Eq. (12), we obtain that two initial states $q_{0}$
and $q_{0}^{'}$ for ${\cal A}$ are equivalent iff they are
$n^2$-equivalent.

\item As in the proof of Theorem 2, given two MO-1QFAs ${\cal
A}_{1}=(Q_{1},\Sigma,q_{0}, \{A_{1}(x)\},F_{1})$ and ${\cal
A}_{2}=(Q_{2},\Sigma,p_{0}, \{A_{2}(x)\},F_{2})$, we let ${\cal
A}=(Q_{1}\cup Q_{2}, \Sigma, \{A_{1}(x)\oplus A_{2}(x):x\in
\Sigma\},F_{1}\cup F_{2})$ (we assume $Q_{1}\cap
Q_{2}=\emptyset$). Then the equivalence between ${\cal A}_{1}$ and
${\cal A}_{2}$ amounts to the equivalence between the initial
states $\rho$ and $\rho^{'}$ with regard to ${\cal A}$, where
$\rho$ and $\rho^{'}$ have corresponding vectors $(\langle
q_{0}|,{\bf 0})$ and $({\bf 0},\langle p_{0}|)$, respectively.

\item In virtue of the   above considerations and the idea in
Subsection 3.1, we describe an algorithm in Figure 2. Analogous to
Theorem 2 and Remark 4, the time-complexity of this algorithm is
$O(m.(n_{1}+n_{2})^{12})$, where $n_1=|Q_1|$, $n_2=|Q_2|$ and
$m=|\Sigma|$.
\end{enumerate}

\begin{center}
 Figure 2.
 Algorithm for the equivalence between MO-1QFAs.
 \fbox{\parbox{15cm}{
 {\small
\begin{quote} {\bf Input:} ${\cal
A}_1=(Q_1,q_{0},\{0,1\},\{A_{1}(x):x=0,1\},F_{1})$, and \\
\hskip 10mm ${\cal
A}_2=(Q_2,p_{0},\{0,1\},\{A_{2}(x):x=0,1\},F_{2})$
\begin{quote}
 Set ${\cal B}$  to be the empty set;\\
 $queue\leftarrow node(\epsilon)$;\\
 {\bf while} $queue$ is not empty {\bf do}\\
 {\bf begin} take an element $F(u)$ from $queue$;
 \begin{quote}
{\bf if} $F(u)$ $\notin span({\cal B})$ {\bf
then} \\
{\bf begin} add vector $F(u)$ to ${\cal B}$;
\begin{quote}
add $F(0u)$ and $F(1u)$ to $queue$;
\end{quote}
{\bf end};
 \end{quote}
 {\bf end};
\end{quote}
\begin{quote}
{\bf if} $\forall B\in {\cal B}$, $(\langle
q_{0}|,{\bf0})B({|q_{0}\rangle,\bf0}) = ({\bf0},\langle
p_{0}|)B({\bf0},|p_{0}\rangle)$
{\bf then} return $(yes)$\\
{\bf else} return (the  string $u$: $(\langle
q_{0}|,{\bf0})B({|q_{0}\rangle,\bf0}) \not= ({\bf0},\langle
p_{0}|)B({\bf0},|p_{0}\rangle)$);
\end{quote}
\end{quote}
}
 }
 }
\end{center}

From the above four steps we have completed the
proof.\qed\\

\begin{Rm}

It is worth indicating that Brodsky and Pippenger [11] and Koshiba
[25] also considered the equivalence problem concerning MO-1QFAs.
Their methods can be described by two steps: (i) firstly using the
bilinearization technique [28] to convert MO-1QFAs to generalized
stochastic finite automata [28]; (ii) secondly determining the
equivalence of generalized stochastic finite automata. Their
difference is regarding step (ii): Koshiba [25] applied the tree
pruning technique [34] to determine the generalized stochastic
systems' equivalence, while Brodsky and Pippenger [11] employed
Paz [30]'s method to do that. Therefore, Koshiba [25] gave a
polynomial-time algorithm for the problem, but Brodsky and
Pippenger [11] did not consider its efficiency. One can find that
our method is different from [25, 11].

\end{Rm}

\subsection* {4.2 A counterexample for the equivalence between MM-1QFAs}
Gruska [22] proposed as an open problem that is it decidable
whether two MM-1QFAs are equivalent. Then   Koshiba [25] tried to
solve the problem. For any MM-1QFA, Koshiba [25] wanted to
construct an equivalent MO-g1QFA (like MO-1QFA but with evolution
matrices not necessarily unitary) and then decided the equivalence
between MO-g1QFAs using the known way on MO-1QFAs. Nevertheless,
we find that the construction technique stated in [25, Theorem 3]
may be not valid, and as a result, the problem is in fact not
solved there. We will give a counterexample to show its
invalidity. In the following, we adopt the definitions of QFAs
stated in [11] where only the right end-marker symbol $\$$ is
considered. So the reader can refer to [11] for the definitions
and we do not detail them here.

First let us  recall the method stated in [25, Theorem 3] for
constructing MO-g1QFAs from MM-1QFAs. Given an MM-1QFA ${\cal
M}=(Q,\Sigma,\{U_\sigma\}_{\sigma\in\Sigma\cup\{\$\}},q_0,
Q_{acc},Q_{rej})$,   an MO-g1QFA ${\cal
M}^{'}=(Q^{'},\Sigma,\{U^{'}_\sigma\}_{\sigma\in\Sigma\cup\{\$\}},q_0,
F)$ is constructed as follows:
\begin{itemize}
    \item  $Q^{'}=Q\cup\{q_\sigma: \sigma\in\Sigma\cup\{\$\}\}\backslash Q_{acc}$, and $F=\{q_\sigma: \sigma\in\Sigma\cup\{\$\}\}$;
    \item  $ U^{'}_\sigma|q\rangle=\dots + \alpha_i|q_i\rangle\dots
    +\alpha_A|q_\sigma\rangle$ when $U_\sigma|q\rangle=\dots + \alpha_i|q_i\rangle\dots +\alpha_A|q_A\rangle$ and $q_A\in Q_{acc}$
  ;
    \item  add the rules:
    $U^{'}_{\sigma}|q_\sigma\rangle=|q_\sigma\rangle$ for all
    $|q_\sigma\rangle\in F$.
    \end{itemize}
Koshiba [25] deemed that the construction technique stated above
can ensure that for any input word, the accepting probability in
${\cal M}$  is preserved in ${\cal M}^{'}$, which will be shown to
be not so.

Now we  turn to the following counterexample provided by us,
showing the invalidity of the above method.
\paragraph{ A counterexample}
Let MM-1QFA ${\cal
M}=(Q,\Sigma,\{U_\sigma\}_{\sigma\in\Sigma\cup\{\$\}},q_0,
Q_{acc},Q_{rej})$, where $Q=\{q_0,q_1,q_{acc},q_{rej}\}$ with the
set of accepting states $Q_{acc}=\{q_{acc}\}$ and the set of
rejecting states $Q_{rej}=\{q_{rej}\}$; $\Sigma=\{a\}$; $q_0$ is
the initial state; $\{U_\sigma\}_{\sigma\in\Sigma\cup\{\$\}}$ are
described below.
\begin{align*}
&U_a(|q_0\rangle)=
\frac{1}{2}|q_0\rangle+\frac{1}{\sqrt{2}}|q_1\rangle+\frac{1}{2}|q_{acc}\rangle,\\
&U_a(|q_1\rangle)=
\frac{1}{2}|q_0\rangle-\frac{1}{\sqrt{2}}|q_1\rangle+\frac{1}{2}|q_{acc}\rangle,\\
&U_\$(|q_0\rangle)=|q_{acc}\rangle,\hspace{2mm}U_\$(|q_1\rangle)=|q_{rej}\rangle.
\end{align*}

Next, we show how this automaton works on the input word $aa\$$.
\begin{enumerate}
\item The automaton starts in $|q_0\rangle$. Then $U_a$ is
applied, giving
$\frac{1}{2}|q_0\rangle+\frac{1}{\sqrt{2}}|q_1\rangle+\frac{1}{2}|q_{acc}\rangle$.
This state is observed with two possible results. With probability
$(\frac{1}{2})^2$, the accepting state is observed and then the
computation terminates. Otherwise, a non-halting state
$\frac{1}{2}|q_0\rangle+\frac{1}{\sqrt{2}}|q_1\rangle$
(unnormalized) is observed
 and then the computation continues.

\item After the second $a$ is fed, the state
$\frac{1}{2}|q_0\rangle+\frac{1}{\sqrt{2}}|q_1\rangle$ is mapped
to
$\frac{1}{2}(\frac{1}{2}+\frac{1}{\sqrt{2}})|q_0\rangle+\frac{1}{\sqrt{2}}(\frac{1}{2}-\frac{1}{\sqrt{2}})
|q_1\rangle+\frac{1}{2}(\frac{1}{2}+\frac{1}{\sqrt{2}})|q_{acc}\rangle$.
This is observed with two possible results. With probability
$[\frac{1}{2}(\frac{1}{2}+\frac{1}{\sqrt{2}})]^2$, the computation
terminates in the accepting state $q_{acc}$. Otherwise, the
computation continues with a new no-halting state
$\frac{1}{2}(\frac{1}{2}+\frac{1}{\sqrt{2}})|q_0\rangle+\frac{1}{\sqrt{2}}(\frac{1}{2}-\frac{1}{\sqrt{2}})
|q_1\rangle$ (unnormalized).

\item After the last symbol $\$$ is fed, the automaton's state
turns to
$\frac{1}{2}(\frac{1}{2}+\frac{1}{\sqrt{2}})|q_{acc}\rangle+\frac{1}{\sqrt{2}}(\frac{1}{2}-\frac{1}{\sqrt{2}})
|q_{rej}\rangle$. This is observed. The computation terminates in
the accepting state $|q_{acc}\rangle$ with probability
$[\frac{1}{2}(\frac{1}{2}+\frac{1}{\sqrt{2}})]^2$ or in the
rejecting state $|q_{rej}\rangle$ with probability
$[\frac{1}{\sqrt{2}}(\frac{1}{2}-\frac{1}{\sqrt{2}})]^2$.
\end{enumerate}

The total accepting probability  is
$(\frac{1}{2})^2+[\frac{1}{2}(\frac{1}{2}+\frac{1}{\sqrt{2}})]^2+[\frac{1}{2}(\frac{1}{2}+\frac{1}{\sqrt{2}})]^2=
\frac{5}{8}+\frac{1}{2\sqrt{2}}$.

Now according to the construction technique [25, Theorem 3] stated
before, we get an MO-g1QFA ${\cal
M}^{'}=(Q^{'},\Sigma,\{U_\sigma^{'}\}_{\sigma\in\Sigma\cup\{\$\}},q_0,F)$
where $Q^{'}=\{q_0,q_1, q_{rej}, q_a, q_\$\}$, $F=\{q_a, q_\$\}$
and $\{U_\sigma^{'}\}_{\sigma\in\Sigma\cup\{\$\}}$ are described
below.
\begin{align*}
&U^{'}_a(|q_0\rangle)=
\frac{1}{2}|q_0\rangle+\frac{1}{\sqrt{2}}|q_1\rangle+\frac{1}{2}|q_a\rangle,\\
&U^{'}_a(|q_1\rangle)=
\frac{1}{2}|q_0\rangle-\frac{1}{\sqrt{2}}|q_1\rangle+\frac{1}{2}|q_a\rangle,\\
&U^{'}_\$(|q_0\rangle)=|q_\$\rangle,\hspace{2mm}U^{'}_\$(|q_1\rangle)=|q_{rej}\rangle,\\
&U^{'}_a(|q_a\rangle)=|q_a\rangle,\hspace{2mm}U^{'}_\$(|q_a\rangle)=|q_a\rangle.
\end{align*}

When the input word is $aa\$$, the automaton works as follows.
Starting from state $|q_0\rangle$, when the first $a$ is fed, the
automaton turns to state
$\frac{1}{2}|q_0\rangle+\frac{1}{\sqrt{2}}|q_1\rangle+\frac{1}{2}|q_a\rangle$.
After the second $a$ is fed, the state is mapped to
$\frac{1}{2}(\frac{1}{2}+\frac{1}{\sqrt{2}})|q_0\rangle+\frac{1}{\sqrt{2}}(\frac{1}{2}-\frac{1}{\sqrt{2}})
|q_1\rangle+[\frac{1}{2}+\frac{1}{2}(\frac{1}{2}+\frac{1}{\sqrt{2}})]|q_a\rangle$.
After the last symbol $\$$ is fed, the state is mapped to
$\frac{1}{2}(\frac{1}{2}+\frac{1}{\sqrt{2}})|q_\$\rangle+\frac{1}{\sqrt{2}}(\frac{1}{2}-\frac{1}{\sqrt{2}})
|q_{rej}\rangle+[\frac{1}{2}+\frac{1}{2}(\frac{1}{2}+\frac{1}{\sqrt{2}})]|q_a\rangle$.

The total accepting probability is
$[\frac{1}{2}(\frac{1}{2}+\frac{1}{\sqrt{2}})]^2+[\frac{1}{2}+\frac{1}{2}(\frac{1}{2}+\frac{1}{\sqrt{2}})]^2=\frac{7}{8}+\frac{1}{\sqrt{2}}$.

Now it turns out  that the accepting probability in the original
MM-1QFA is not preserved in the constructed machine as expected in
[25]. Therefore, the invalidity of the method [25, Theorem 3] has
been shown.

\begin{Rm}
(1) The essential reason for the invalidity of the way in [25] is
that the accepting state set $F$ in ${\cal M}^{'}$  does not
cumulate the  accepting probabilities in the original MM-1QFA.
Instead, it accumulates just the accepting amplitudes. In
addition, we know that in general, $|a|^2+|b|^2\neq|a+b|^2$.
Therefore, the way in [25] leads to invalidity. (2)  Due to the
complex accepting behavior of MM-1QFAs, it is likely no longer
valid to decide the equivalence between MM-1QFAs as we did for
MO-1QFAs. To our knowledge, so far there seems to be no existing
valid solution to this problem. Therefore, the equivalence between
MM-1QFAs is worth considering further.
\end{Rm}

\section*{5. Concluding remarks}

 In this paper, based on the results in [27, 31], we presented a
polynomial-time algorithm $(O(m.(n_{1}+n_{2})^{12}))$ for
determining the equivalence between two QSMs with $n_{1}$ and
$n_{2}$ states, respectively, and, if they are not equivalent,
this algorithm will produce an input-output pair with length not
more than $(n_{1}+n_{2})^{2}$. Furthermore, by using the way of
Moore and Crutchfield [28], we obtained that two QSMs ${\cal
M}_{1}$ and ${\cal M}_{2}$ are equivalent iff they are
$(n_1^2+n_2^2-1)$-equivalent, which improves the result in [27].

We also proved that two MO-1QFAs ${\cal A}_{1}$ and ${\cal A}_{2}$
that have $n_{1}$ and $n_{2}$ states, respectively, and the same
input alphabet $\Sigma$ with $|\Sigma|=m$, are equivalent if, and
only if they are $(n_{1}+n_{2})^{2}$--equivalent. In terms of the
idea of the algorithm for QSMs, we further provided a
polynomial-time algorithm $(O(m.(n_{1}+n_{2})^{12}))$ for the
equivalence between ${\cal A}_{1}$ and ${\cal A}_{2}$.

In addition, considering the problem of deciding the equivalence
between MM-1QFAs, we provided a counterexample showing that the
method stated in [25] to solve the problem may be not valid, and
therefore
 the problem is left open again.

The further problems are regarding the minimization of states for
QSMs [20,31]. As well, the equivalence concerning 2QFAs [26] is
worthy of consideration.

\end{document}